\documentclass[preprint,12pt,authoryear]{elsarticle}
\pdfoutput=1 

\usepackage{amssymb}
\usepackage{amsmath}
\usepackage{booktabs}
\usepackage{array}
\usepackage{longtable}
\usepackage{tabularx}
\usepackage{xltabular}
\usepackage{makecell}
\usepackage{enumitem}
\usepackage[protrusion=true,expansion=false]{microtype}
\usepackage{xurl}

\newcolumntype{Y}{>{\raggedright\arraybackslash}X}
\journal{Information and Software Technology}

\begin{document}

\begin{frontmatter}

\title{Domain-Validity-Gated Metamorphic Testing of Scientific ML Surrogates}

\author[inst1,inst2,inst3]{Meng Li\corref{cor1}}
\ead{mlemon@usc.edu.cn}
\author[inst1,inst2,inst3]{Xiaohua Yang}
\ead{xiaohua1963@foxmail.com}
\author[inst1,inst2,inst3]{Jie Liu}
\author[inst1,inst2,inst3]{Shiyu Yan}
\cortext[cor1]{Corresponding author.}
\affiliation[inst1]{organization={School of Computing, University of South China},
  city={Hengyang}, country={China}}
\affiliation[inst2]{organization={Hunan Engineering Research Center of Software Evaluation and Testing for Intellectual Equipment},
  city={Hengyang}, country={China}}
\affiliation[inst3]{organization={CNNC Key Laboratory on High Trusted Computing},
  city={Hengyang}, country={China}}

\begin{abstract}
\textbf{Context:}
Scientific machine-learning (SciML) surrogates approximate expensive physical simulations, but exact expected outputs for arbitrary inputs are unavailable (the classical oracle problem). Metamorphic testing addresses this through relations across executions, yet a candidate relation is not automatically valid: its preconditions, output mapping, and the numerical floor of the scoring operator determine whether a violation is meaningful.

\textbf{Objective:}
We investigate how candidate metamorphic relations (MRs) can be screened for \emph{domain validity} and operationalized as executable, oracle-free test assets for SciML surrogates, making the step from candidate to executable asset auditable and explicit about the physical, numerical, and software conditions each relation requires.

\textbf{Method:}
We propose (i) a domain-validity rubric admitting a candidate only when its tolerance dominates the operator's numerical floor and its preconditions hold; (ii) an MR-card / executable-asset format recording source cases, transformations, mappings, metrics, tolerances, exclusion rules, and relation-level verdicts; and (iii) a case-study protocol on MeshGraphNets cylinder-flow surrogates; a claim ledger binds every result to a tracked artifact.

\textbf{Results:}
On a MeshGraphNets checkpoint, node permutation holds to machine precision; mirror-y is a bounded OOD-stress finding, not an admitted exact symmetry; absolute conservation remains deferred while a reference-relative non-regression guard passes. The same readings hold across held-out trajectories, the checkpoint roster, three further cylinder-flow architectures, and PhysicsNeMo. On a second CFD task (compressible airfoil) the predicate instead rejects incompressible continuity on physical grounds, showing it reasons about domain validity rather than running a fixed checklist. The divergence operator follows the expected order trend; on a second PDE family, FNO Burgers/heat surrogates run full admit/reject/execute verdicts; and a unified fault catalogue reports detector coverage and effect sizes.

\textbf{Conclusion:}
Evidence spanning two CFD tasks and a second PDE family supports a validity-aware bridge from candidate MRs to auditable SciML test assets, separating model-level violations from out-of-domain applications. Broader generalization is future work.
\end{abstract}

\begin{highlights}
\item A domain-validity rubric screens candidate MRs for SciML surrogate testing.
\item Each MR tolerance is gated by the operator's own $O(h)$ numerical floor.
\item MR cards convert retained candidates into auditable executable test assets.
\item A fail-closed claim ledger ties every reported number to a tracked artifact.
\item K=6 rosters expose where MR detectors are structurally insensitive.
\end{highlights}


\end{frontmatter}

\section{Introduction}
\label{sec:introduction}

Scientific computing increasingly uses learned surrogates to reduce the cost of repeated numerical simulation. Mesh-based neural simulators such as MeshGraphNets learn to predict physical fields on irregular meshes \citep{pfaff2021meshgraphnets}. These models approximate high-cost solvers but introduce a familiar software testing problem: for many candidate inputs there is no cheap exact expected output against which the program can be checked---the test oracle problem \citep{barr2015oracle}.

The common response is to evaluate on held-out trajectories and report rollout error. This is necessary but incomplete: a model may have acceptable average error while failing to preserve relations that should hold under physically meaningful transformations. For a user deploying a SciML surrogate outside a narrow validation set, the question is not only ``How accurate is the model?'' but also ``Under which transformations does the system stop respecting relations it should respect?''

Metamorphic testing (MT) addresses the oracle problem by checking relations among multiple executions \citep{chen1998metamorphic,segura2016survey}. Conservation laws, symmetry relations, nondimensional similarity, and continuity conditions all suggest possible relations among executions \citep{kanewala2019scientific,lin2020exploratory,olsen2019simulation}. The difficulty is deciding which candidates are valid and executable for a particular SciML program: for cylinder flow, mirror symmetry depends on geometry and boundary labels; a divergence check depends on a discrete operator, mesh weights, and boundary treatment. Treating such transformations as automatically valid MRs hides the assumptions that determine whether a violation is meaningful.

The remaining problem addressed here is therefore narrower: how candidate MRs are screened for domain validity, turned into executable test assets, and reported through relation-level verdicts that distinguish SUT inconsistency from out-of-relation-domain cases and numerical tolerance effects.

This paper treats MR identification for SciML as a validity-gated testing problem. Physical knowledge, expert reasoning, LLM-assisted candidate lists, and NOETHER-style pattern organization \citep{zhao2026noether} can all suggest candidate relations, but a candidate is not yet an executable MR: it must first state the physical basis, transformation preconditions, boundary-condition compatibility, output mapping, metric, tolerance rationale, exclusion rule, and verdict interpretation.

Two ideas organize this treatment. First, a candidate is \emph{admissible} only when numerically decidable: its verdict tolerance must dominate the intrinsic error floor of the measuring operator---machine precision for an exact relation, the mapping floor for an approximate geometric relation, or the discretization floor for a continuity relation. Second, a relation-level verdict is read in two dimensions: violation magnitude and domain-violation magnitude, so a model inconsistency is not confused with a relation applied outside its domain. We refer to this as a domain-admissibility-gated, relation-indexed approach to SciML OOD validation. The primary case study is MeshGraphNets-family cylinder flow, with the same predicate additionally exercised on a second CFD task (compressible airfoil) and a second PDE family (FNO/PINN on Burgers and heat); full cross-SUT rate estimation remains blocked.

\subsection{Research questions}
\label{subsec:rqs}

The main research question is:

\textbf{RQ0.} How can candidate metamorphic relations for scientific machine-learning systems be screened for domain validity and converted into executable oracle-free test assets without relying on exact per-sample expected outputs?

We decompose this into four questions:
\begin{description}[leftmargin=!,labelwidth=2.4cm]
\item[RQ1 -- Validity.] How can a domain-validity rubric distinguish physically meaningful candidate MRs from transformations that are executable but invalid, underspecified, or outside the relation's domain?
\item[RQ2 -- Operationalization.] How can retained candidates be represented as MR cards and executable assets with source cases, follow-up transformations, metrics, thresholds, exclusions, and verdict rules?
\item[RQ3 -- Verdict.] How can relation-level verdicts distinguish pass, fail, skip, out-of-relation-domain, numerical tolerance issue, and inconclusive outcomes?
\item[RQ4 -- Case-study evidence.] In a MeshGraphNets-family cylinder-flow case study, what evidence does the rubric-gated asset workflow add relative to rollout-accuracy diagnostics, secondary LLM/generic candidate baselines, and external witness evidence?
\end{description}

\subsection{Contributions}
\label{subsec:contributions}

This paper makes three scoped contributions, each repositioned narrowly with respect to the closest prior identified in Section~\ref{subsec:physics-mt}. The positive claim is methodological: the paper contributes a measurement-floor admissibility gate, a typed domain-inadmissibility verdict, and a seeded-fault diagnostic stress test: the MR-class-to-fault-class diagnostic is used as stress-test evidence, not a validated localization model.

\textbf{Measurement-floor admissibility gate.} A domain-validity rubric and MR-card executable-asset workflow screen candidate MRs and convert retained ones into auditable executable test assets recording physical basis, transformation preconditions, output mapping, metric, tolerance, and exclusion rules. We ground the tolerance floor in the \emph{intrinsic error floor of the discrete measurement operator} (theoretically $O(h)$ for a P1 divergence operator on a triangular mesh); for the concrete deployed-scale mesh this floor is closed-form---a leading-order predictor matches the measured floor to within 0.5\% and a rigorous a-priori bound dominates it (Sec.~\ref{subsec:operator-floor-sweep})---with a general unstructured-mesh bound left to future work. To our knowledge, grounding an MR tolerance in the measurement operator's own error floor is new in the SciML setting.

\textbf{Typed domain-inadmissibility verdict.} A relation-level verdict and ledger scheme reads verdicts in two dimensions (relation-violation against domain-violation magnitude). This instantiates, for physics-governed SciML, the constraint-architecture pattern of Duque-Torres et al.~\citep{duqueTorres2023bugornot,duqueTorres2023completePipeline} and MetaTrimmer~\citep{duqueTorres2023metaTrimmer}, adding a \emph{typed classification} of domain-inadmissibility sub-dimensions drawn from PDE preconditions, geometry, boundary conditions, and operator admissibility.

\textbf{Seeded-fault diagnostic stress test.} The paper's own MRs are used as fault detectors against an independently re-implemented 10-mutant seeded-fault catalogue, reporting a by-class stress-test pattern: continuity responds to boundary-condition/normalization faults; symmetry responds to physical-channel/adjacency faults; node-permutation responds to none because the tested faults preserve relabeling invariance. This is suggestive evidence for a typed MR-class-to-fault-class diagnostic, but it remains a stress test over seeded faults rather than a localization model.

To keep the contribution boundary easy to follow: the three devices above are the claims; the within-SUT pilots, replication, operator-floor, and fault-detection subsections carry their primary evidence; the cross-family subsections show the predicate transferring across model families; everything else (baselines, witness audits) is context, not claim.

We evaluate these contributions on one task with deliberately wide architectural coverage: a trained MeshGraphNets checkpoint, its K=6 roster over three official held-out trajectories, same-domain S4/S5 wider/deeper variants, a non-message-passing PointMLP network, and the NVIDIA PhysicsNeMo production implementation evaluated on official DeepMind cylinder\_flow data. Cross-family PINN and FNO executions check that the same admissibility predicate transfers outside mesh surrogates, and rollout-accuracy, exact mirror-y, Minimum-MR-SubSet rerun, and generic-MR / LLM-candidate comparators complete the design.

\section{Background and Related Work}
\label{sec:related}

\subsection{Mesh-based neural simulation}
\label{subsec:mesh-simulation}

Mesh-based neural simulators learn dynamics on graph or mesh representations of physical systems. MeshGraphNets encodes mesh nodes and edges, propagates information by message passing, and predicts future physical fields through autoregressive rollout \citep{pfaff2021meshgraphnets}. The cylinder-flow benchmark exposes the distinction between data-driven accuracy and physical relation preservation. In this paper, MeshGraphNets-family systems are treated as software systems under test: we ask how outputs behave under controlled input transformations and whether necessary relations hold within explicit tolerances.

\subsection{Metamorphic testing and the oracle problem}
\label{subsec:mt-oracle}

Metamorphic testing (MT) addresses programs lacking a per-input oracle \citep{chen1998metamorphic,segura2016survey,chen2018mtSurvey}: instead of checking each output against an expected value, MT checks a relation among outputs of a source and follow-up inputs. This framing has been applied to scientific software, simulations, and ML systems \citep{chen2011ml,kanewala2019scientific,olsen2019simulation}. For SciML surrogates the oracle problem is especially visible in OOD settings, where a trusted solver may be too expensive to run for every transformed case. SciML MRs cannot be treated as generic input perturbations: every transformation must respect governing equations, boundary conditions, mesh representation, and numerical tolerance, making MR correctness a central validity question.

\subsection{MR identification for scientific and simulation software}
\label{subsec:mr-identification}

MRs can be elicited from monotonicity, conservation, scaling, and domain-specific expectations \citep{kanewala2019scientific,lin2020exploratory,olsen2019simulation,raunak2021continuum}. Predictive MR identification has been pursued with graph-kernel learning over scientific source code \citep{kanewala2016graphkernel}, treating MR discovery itself as a learning task. Data-driven search can also discover candidate transformations in complex physical software \citep{hiremath2021ocean}, and design assumptions can be encoded as MRs for CPS testing under control--plant invariants \citep{mandrioli2025cps}. The gap is that candidate identification is insufficient for SciML: without a domain-of-validity record a failed test is ambiguous: the relation may have been invalid under the chosen boundary conditions, the transformation may have left the physical regime, the tolerance may not have been numerically justified, or the SUT may have violated a genuine constraint. This paper makes that distinction explicit.

\subsection{Physics-based MT for learned scientific simulators}
\label{subsec:physics-mt}

Three contemporary works are the closest prior. \textbf{Reichert et al.~(2024)}~\citep{reichert2024hess} applied physics-derived MRs to a trained LSTM hydrologic surrogate, stratifying pass/fail by basin elevation and informally filtering basins where forcing uncertainty dominated the response. We formalise these practices as an explicit admissibility predicate and a two-dimensional verdict, and extend them to mesh-based neural fluid surrogates whose MR validity depends on geometry and discrete operators.

\textbf{Eniser et al.~(2022)}~\citep{eniser2022relaxations} introduced \emph{relaxations}---numerical tolerances embedded in MR oracles---for stochastic RL policy testing. We extend this calibrated-tolerance principle to deterministic surrogates by grounding the tolerance floor in the intrinsic error floor of the discrete measurement operator (theoretically $O(h)$ for a P1 divergence operator); for the concrete deployed-scale mesh the floor is given in closed form (a leading-order predictor within 0.5\% plus a rigorous a-priori bound; Sec.~\ref{subsec:operator-floor-sweep}), with a general unstructured-mesh bound left to future work.

\textbf{The 2023 violation-attribution cluster}---Duque-Torres et al.~\citep{duqueTorres2023bugornot,duqueTorres2023completePipeline} and MetaTrimmer~\citep{duqueTorres2023metaTrimmer}---addressed the bug-vs-inapplicability separation with explicit MR constraints and automated constraint derivation. Our two-dimensional verdict instantiates this pattern for physics-governed SciML, adding a \emph{typed classification of domain-inadmissibility sub-dimensions} drawn from PDE preconditions, geometry, boundary conditions, and operator admissibility. The element none of these three works addresses is the seeded-fault MR-as-detector with by-class localization of Section~\ref{subsec:within-sut-pilot}.

\paragraph{Closest-prior positioning.}
\label{tab:closest-prior-positioning}
Reichert et al.: physics-MR applicability filtering; Eniser et al.: relaxations; Duque-Torres / MetaTrimmer: bug-vs-inapplicability constraints. What this paper adds: measurement-floor admissibility, typed domain-inadmissibility, and MR-class-to-fault-class diagnostics for SciML.

PINN training itself exhibits gradient-flow pathologies that bias trained models away from the governing residual \citep{wang2021gradflow,krishnapriyan2021failure}, sharpening the case for relation-level validation that does not assume a converged surrogate.

\subsection{SciML V\&V, residuals, UQ, and failure modes}
\label{subsec:sciml-vv}

SciML reliability research has developed complementary tools: physics residuals, uncertainty quantification, conformal prediction, certified error bounds, and failure-mode analysis \citep{raissi2019pinn,karniadakis2021piml,li2021fno,krishnapriyan2021failure,gopakumar2025calibrated}. A residual or equivariance error becomes part of an executable MR only when paired with a source case, a follow-up transformation, a tolerance rule, and a verdict interpretation. We use SciML diagnostics as possible relation measurements; MT supplies the multi-execution oracle-free structure.

\subsection{Hybrid ML-solver trust regions}
\label{subsec:hybrid-trust}

Hybrid ML-solver frameworks use residual thresholds or error estimates to decide when a learned component should be trusted \citep{baral2025xrepit,wang2025deeponetfe}. Such systems operationalize runtime trust regions, whereas the present study acts offline: physically derived transformations are applied before deployment to estimate where relation violations occur.

The distinction from residual- and uncertainty-based trust estimation is deliberate. UQ, conformal prediction, and residual-threshold trust regions locate unreliable behavior passively from observed inputs. The present method acts in relation space: it applies a controlled transformation and reports which necessary relation breaks under it, indexed by that transformation. The two-dimensional verdict separates a model-level violation from an out-of-domain application, a distinction a scalar accuracy or residual magnitude does not by itself provide. Section~\ref{subsec:within-sut-pilot} gives the concrete instance: a surrogate accurate in-distribution (median one-step relative L2 0.0216) still violates mirror-y equivariance by roughly an order of magnitude, so accuracy does not bound the relation violation.

\subsection{What is new and what is not new}
\label{subsec:new-not-new}

Metamorphic testing, MR identification, scientific-software MT, residual diagnostics, uncertainty quantification, LLM candidate generation, and NOETHER-style candidate organization are established or emerging sources of testing ideas. The narrower claim is that SciML MR identification should be treated as a domain-validity problem: a candidate relation becomes useful only after its physical basis, transformation preconditions, output mapping, tolerance, exclusion rule, executable artifact, and relation-level verdict are recorded. Physics-based MRs for learned field predictors are an active area; the verified record sharpens the novelty statement toward validity-gated execution rather than first-use positioning.

What is new is the evidence-gated conversion from candidate relation to executable SciML MR asset---to our knowledge the first end-to-end validity-gated metamorphic-testing pipeline for physics-governed SciML in which every stage, from candidate screening through verdict typing to fault-class diagnosis, is operationalized and executed rather than proposed. Each closest prior contributes one ingredient---an informal admissibility filter, a calibrated tolerance, a binary applicability pre-filter---but none combines them, grounds the tolerance in the measurement operator's own characterizable error floor, types the inadmissibility verdict, or maps MR failures back to fault classes. The structural novelty is two organizing devices: an admissibility gate tying a relation's tolerance to the numerical error floor of its own measurement, and a two-dimensional verdict separating a model violation from an out-of-domain application; the same predicate executes unchanged across MeshGraphNet, PointMLP, PhysicsNeMo, PINN, and FNO subjects. The admissibility gate is fully operational in the case study; the verdict's domain-violation axis is operationalized per relation as a continuous score from committed precondition measurements (Section~\ref{subsec:verdicts}); calibrating these scores across MR classes is still future work. A third empirically distinct element, examined in Section~\ref{subsec:within-sut-pilot}, is the seeded-fault diagnostic stress test: the MRs show a by-class response pattern (continuity to boundary/scale faults, symmetry to physical-channel/adjacency faults), an element none of the three closest prior works addresses.

\section{Method}
\label{sec:method}

\subsection{Overview}
\label{subsec:method-overview}

\begin{figure}[!htbp]
\centering
\includegraphics[width=0.85\linewidth]{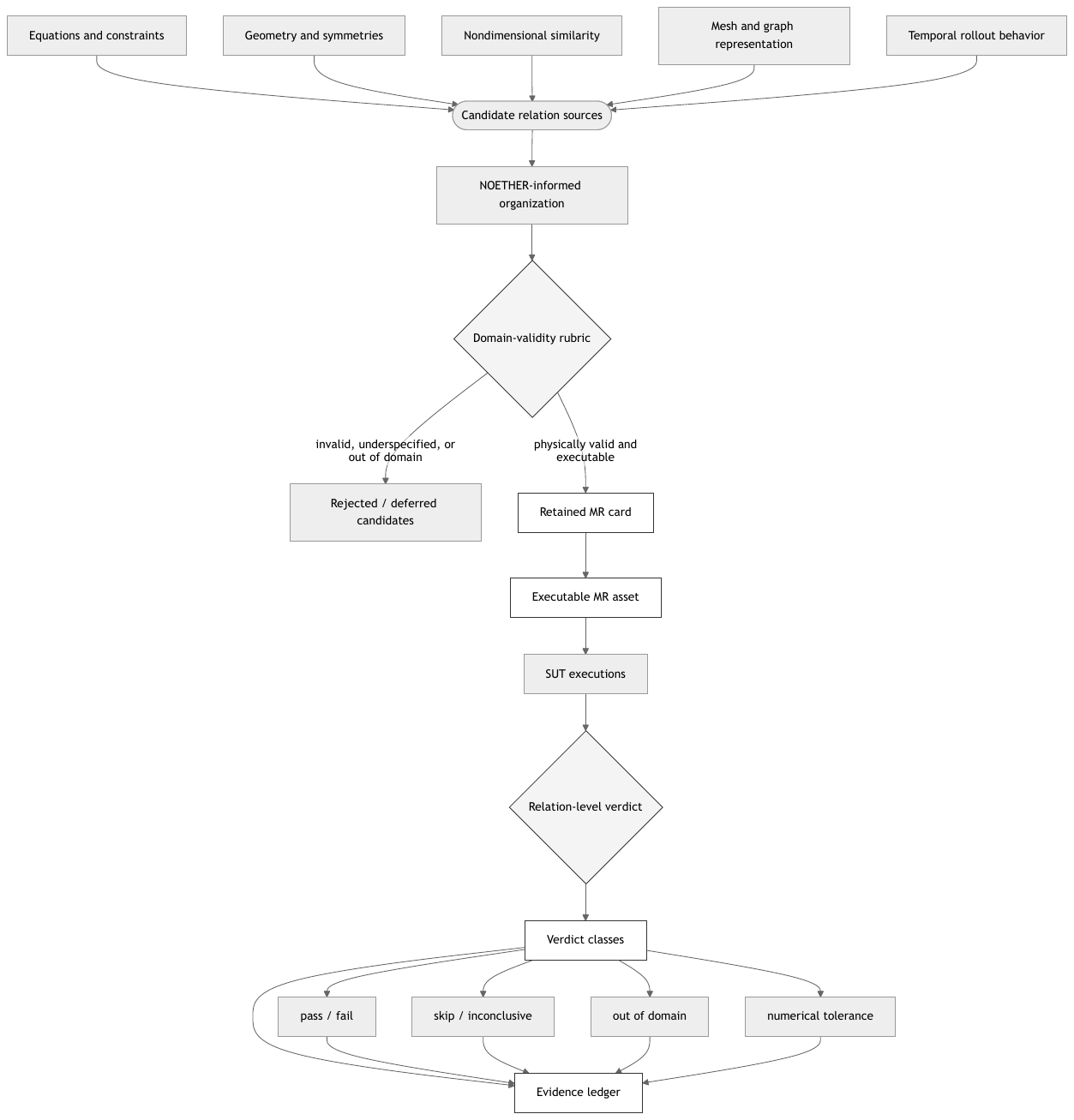}
\caption{Validity-gated workflow for converting physically motivated candidate relations into executable, oracle-free MR assets and relation-level verdicts. The rubric and the verdict are explicit gates (diamonds); rejected candidates and out-of-relation-domain cases are recorded as evidence rather than discarded silently.}
\label{fig:workflow}
\end{figure}

The proposed method is a five-stage workflow: (1)~identify candidate relation sources; (2)~organize candidates using declared source categories; (3)~screen with a domain-validity rubric; (4)~convert retained relations into executable MR assets; (5)~execute assets and record relation-level verdicts. Candidate generation is deliberately separated from validity judgment: NOETHER-style patterns may help organize candidates but do not certify that a relation is physically valid for a particular SUT. Figure~\ref{fig:workflow} summarizes the workflow.

\subsection{Candidate relation sources}
\label{subsec:candidate-sources}

Candidate MRs are read off the algebraic properties of the governing and discrete operators, each located in one of three levels (physical-model, computational-model, code-model): \textbf{equivariance} (reflections/rigid motions give physical-model symmetry MRs under compatible geometry; node/face permutation gives code-model representation contracts); \textbf{conservation} (discrete-divergence/boundary-flux relations, decidable only when the operator floor permits, Sec.~\ref{subsec:operator-floor-sweep}); \textbf{homogeneity/scaling} (Reynolds/Strouhal, within regime); \textbf{composition} (rollout determinism/prefix-consistency---implementation checks, not physical laws); and \textbf{cross-implementation comparison} (only when units, meshes, rollout horizons, and checkpoints are comparable). The same property can hold at the physical-model level yet fail at the computational or code level---an asymmetric mesh breaks reflection-equivariance, an uncalibratable operator floor blocks conservation decidability---so the admissibility gate locates where each property survives and the typed verdict records it.

\subsection{Domain-validity rubric}
\label{subsec:rubric}

A candidate relation is an \emph{admissible MR} when four conditions hold together: (i) it has a physical or software basis; (ii) its transformation preconditions are satisfied; (iii) boundary conditions and output mapping remain compatible after the transformation; and (iv) it is numerically decidable---the verdict tolerance dominates the intrinsic error floor of the measuring operator. None of the four is optional: each is a gate that can reject or defer a candidate. The six rubric criteria in Table~\ref{tab:rubric} are the auditable form of these four conditions.

\begin{table}[t]
\centering
\caption{Domain-validity rubric for candidate MRs.}
\label{tab:rubric}
\begin{tabularx}{\textwidth}{>{\raggedright\arraybackslash}p{0.24\textwidth}X}
\toprule
\textbf{Criterion} & \textbf{Question asked before retention} \\
\midrule
Physical basis & What equation, boundary condition, nondimensional law, representation property, numerical assumption, or implementation contract justifies the relation? \\
Transformation preconditions & What exactly changes from source to follow-up, and what must be preserved? \\
Boundary-condition compatibility & Does the transformation preserve the intended physical meaning of the domain and boundaries? \\
Output mapping & Can the expected output relation be measured from available SUT outputs? \\
Metric and tolerance & What metric is used, and where does the tolerance come from? \\
Failure diagnosability & What can a violation plausibly mean, and what can it not mean? \\
\bottomrule
\end{tabularx}
\end{table}

The rubric yields one of four decisions: retained as executable MR; retained as OOD stress relation; rejected; or deferred pending missing evidence. The defer and OOD-stress decisions are the two soft-fail paths: deferred when condition (iv) cannot yet be established (e.g.\ an absolute discrete-divergence relation whose reference field already carries non-negligible divergence), and downgraded to OOD-stress when conditions (ii)--(iii) hold only approximately.

\subsection{MR card and executable asset format}
\label{subsec:mr-card}

A retained MR is represented as an MR card recording: identifier, source category, physical or software basis, source-case schema, follow-up transformation, preconditions, boundary-condition compatibility, output mapping, metric, tolerance and provenance, expected verdict classes, exclusion rules, artifact schema, and fault interpretation. Initial card skeletons for the cylinder-flow study are listed with their verdicts in Tables~\ref{tab:claim-evidence} and~\ref{tab:mr-card-verdict}.

\begin{figure}[!htbp]
\centering
\includegraphics[width=0.85\linewidth]{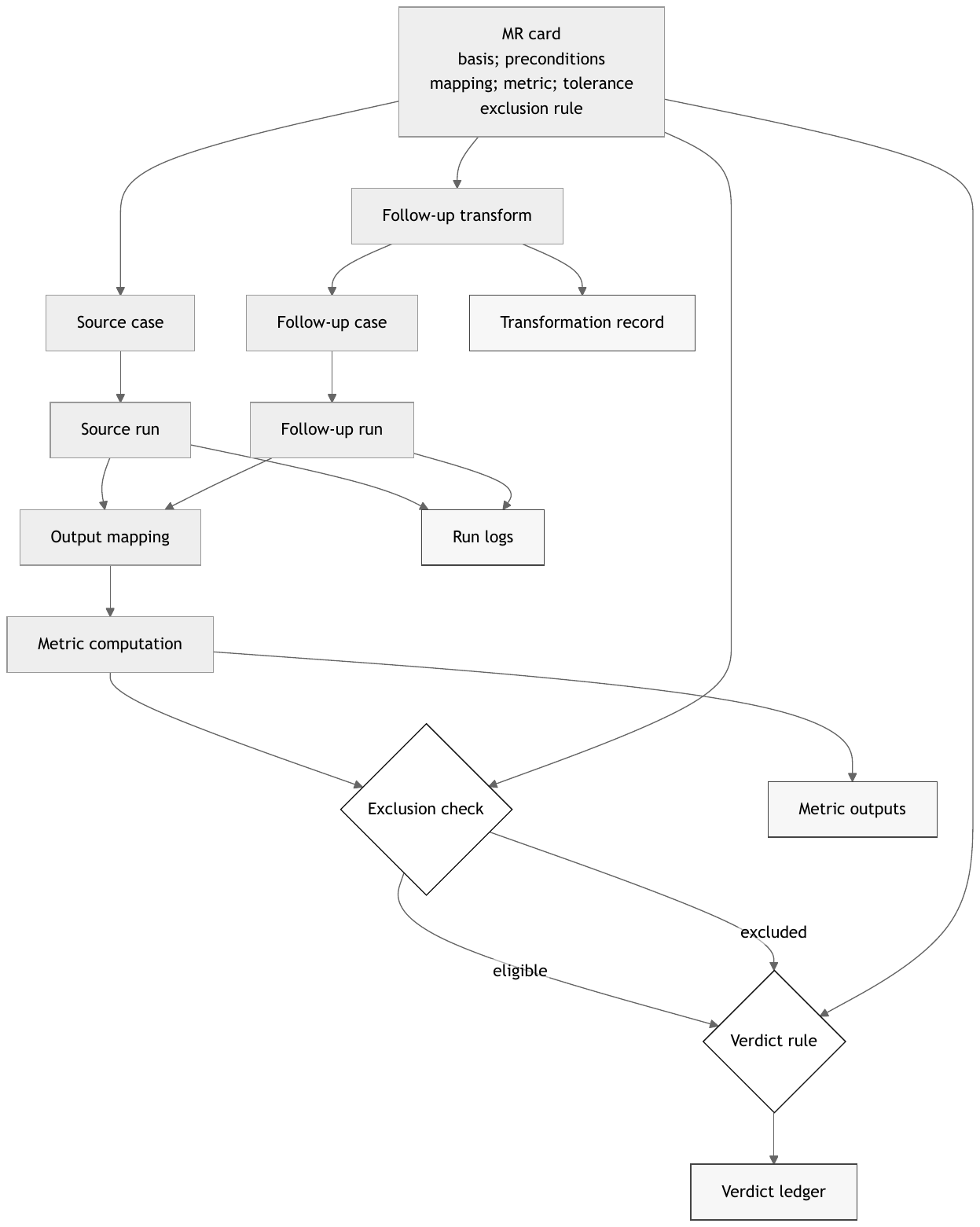}
\caption{Executable MR asset and verdict data flow. The MR card supplies the basis, preconditions, mapping, metric, tolerance, and exclusion rule; the source and follow-up runs feed the output mapping and metric; the exclusion check and verdict rule emit a relation-level verdict to the ledger together with metric outputs, run logs, and the transformation record as artifacts.}
\label{fig:asset-flow}
\end{figure}

Executable assets are generated from MR cards; each includes transformation code, runner configuration, metric computation, verdict logic, and artifact recording. Figure~\ref{fig:asset-flow} shows the resulting data flow.

\subsection{Worked examples: from candidate relation to executable MR card}
\label{subsec:worked-examples}

These are design-time screening decisions, not empirical findings, and do not report whether any SUT passes or fails. Table~\ref{tab:worked-decisions} summarizes three such decisions, stressing different parts of the method: representation semantics, physical symmetry under boundary conditions, and a constraint measurement whose validity depends on a discrete operator and tolerance provenance.

\begin{table}[t]
\centering
\caption{Worked screening decisions for three candidate MRs.}
\label{tab:worked-decisions}
\small
\setlength{\tabcolsep}{3pt}
\begin{tabularx}{\linewidth}{@{}>{\raggedright\arraybackslash}p{0.17\linewidth}>{\raggedright\arraybackslash}p{0.24\linewidth}>{\raggedright\arraybackslash}p{0.20\linewidth}Y@{}}
\toprule
\textbf{Candidate} & \textbf{Main validity issue} & \textbf{Screening decision} & \textbf{Reason} \\
\midrule
Node permutation equivariance & Graph relabeling must preserve the same physical mesh and fields & Retain as executable representation MR & The transformation changes representation order but not the physical case, provided the adapter and inverse output mapping are deterministic. \\
Mirror-y equivariance & Geometry, boundary labels, and vector components must transform consistently & Retain only under boundary-compatible conditions & The relation is physically meaningful for symmetric cylinder-flow cases, but invalid if the transformation changes inlet, outlet, wall, or obstacle semantics. \\
Discrete divergence boundedness & The metric depends on the discrete divergence operator, mesh weights, boundary treatment, and tolerance & Defer or retain as qualified continuity-constraint MR & The physical basis is strong for incompressible flow, but the executable verdict is meaningful only when the numerical operator and threshold are justified. \\
\bottomrule
\end{tabularx}
\end{table}

\paragraph{MR-1: node permutation equivariance.}
A permutation $\pi$ is applied to node order and corresponding edge indices without changing coordinates, fields, or boundary labels. Equivariance requires that the output for the permuted case equals $\pi(Y)$ up to numerical tolerance; the output mapping applies $\pi^{-1}$ before comparison. This is a representation-level contract: tolerance is derived from deterministic repeat runs or machine precision; a failure points to graph adapter, batching, or message-passing implementation issues, not a Navier--Stokes violation.

\paragraph{MR-2: mirror-y equivariance.}
Coordinates $(x,y)$ map to $(x,-y)$; the transverse velocity component changes sign; boundary labels transform with the mesh. Retained only when cylinder, inlet, outlet, and wall labels remain semantically compatible after reflection; rejected if the transformation swaps non-equivalent boundaries or leaves vector components unmapped. A violation may indicate a SUT symmetry inconsistency or that the MR was applied outside its domain; the relation-level verdict must preserve that distinction.

\paragraph{MR-3: discrete divergence boundedness.}
Predicted velocity fields should remain within a bounded discrete divergence tolerance for eligible interior cells. Retained only after the discrete operator, mesh weights, boundary treatment, and tolerance provenance are specified; the exclusion rule skips cases where boundary treatment dominates the metric or where the flow is outside the incompressibility assumption. A failure is a continuity-constraint violation only after operator and tolerance evidence rule out a numerical artifact.

These examples show how the rubric changes the status of candidate relations: node permutation becomes a retained representation MR, mirror-y a conditional physical-symmetry MR, and divergence a qualified continuity-constraint MR whose verdict depends on numerical instrumentation.

\subsection{Relation-level verdicts}
\label{subsec:verdicts}

An MR execution can produce: \textbf{pass} (relation holds within tolerance); \textbf{fail} (violated within valid domain); \textbf{skip} (precondition missing); \textbf{out-of-relation-domain} (transformation produced a case outside the validity domain); \textbf{numerical-tolerance issue} (dominated by threshold/resolution uncertainty); or \textbf{inconclusive} (insufficient artifact).

These verdicts form a two-dimensional space. One axis is relation-violation magnitude (V/floor ratio); the other is domain-violation magnitude (how far the transformed case lies outside the validity domain). Low domain-violation with high relation-violation is the only region that may be read as SUT inconsistency; high domain-violation is out-of-relation-domain; a violation within the error floor is a numerical-tolerance issue. This decomposition keeps a model-level violation from being confused with a relation applied outside its domain.

\begin{figure}[!htbp]
\centering
\includegraphics[width=0.92\linewidth]{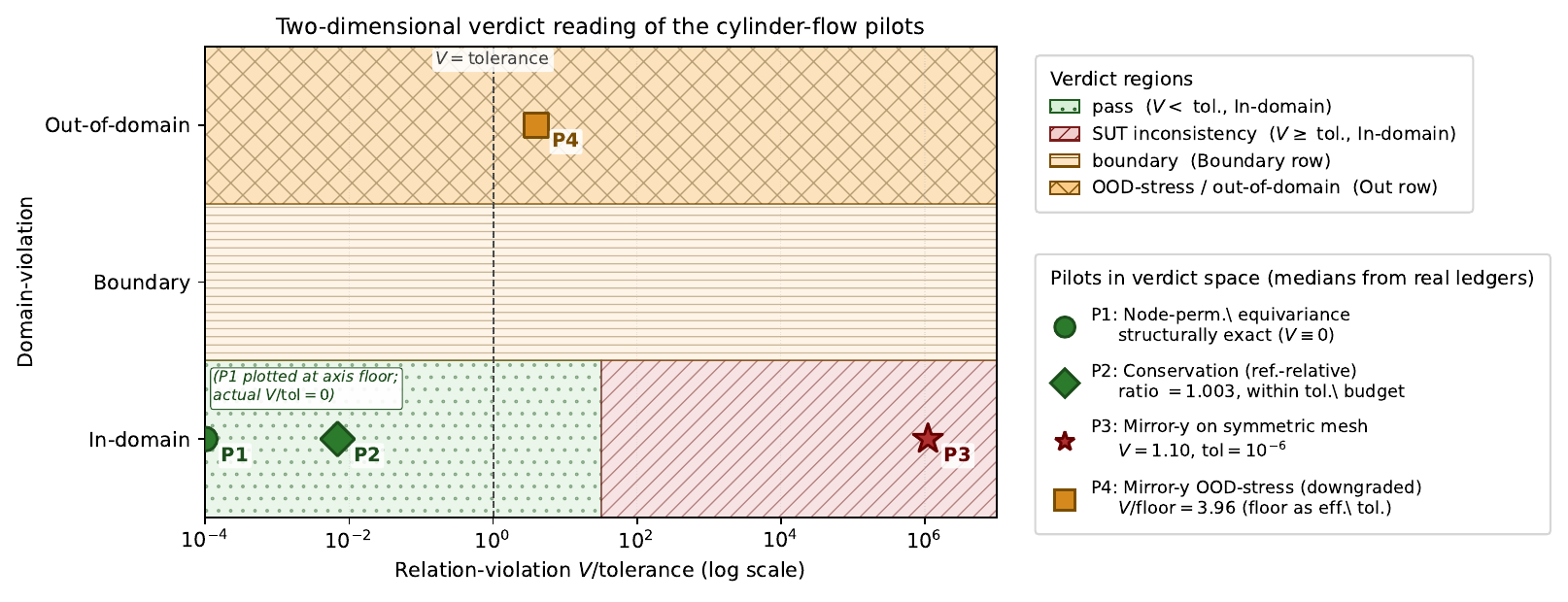}
\caption{Two-dimensional verdict reading of the cylinder-flow pilots. Horizontal axis: relation-violation $V/\mathrm{tolerance}$ on a log scale, where tolerance is the per-MR verdict threshold (machine-precision floor for an exact relation; the reference-relative regression threshold for conservation; the mapping-error floor when an approximate relation has been downgraded to an OOD-stress probe, as for P4). Vertical axis: domain-violation as a three-level categorical bin (In-domain, Boundary, Out-of-domain); the bins are backed by the per-relation continuous scores of Sec.~\ref{subsec:verdicts}, which are not mutually calibrated across relations. The dashed line at $V/\mathrm{tolerance} = 1$ is the pass/fail border for an admissible relation. Coordinates (medians from the metric ledgers; full numbers in Sec.~\ref{subsec:within-sut-pilot} and Table~\ref{tab:mr-card-verdict}): P1 node-permutation equivariance $V = 0$ exactly (structurally exact, plotted at the log-axis floor); P2 conservation (reference-relative) ratio $= 1.005$, so $V/\mathrm{tolerance} \approx 0.01$, two orders of magnitude above P1 yet still well inside pass; P3 mirror-y on the synthetic symmetric mesh, where the rubric admits the exact relation, $V = 1.10$ against tolerance $10^{-6}$, $V/\mathrm{tolerance} \approx 1.1\times 10^{6}$: the SUT-inconsistency region, which is what removes the out-of-relation-domain objection to the mirror-y finding; P4 mirror-y OOD-stress (downgraded), $V/\mathrm{floor} = 3.96$ with the mapping-error floor as effective tolerance, sitting in the Out-of-domain row. Absolute conservation is not plotted: its admissibility predicate (Sec.~\ref{subsec:rubric}) fails because the absolute tolerance does not dominate the P1 operator floor calibrated in Sec.~\ref{subsec:operator-floor-sweep}, so the rubric never admits it into the verdict space.}
\label{fig:verdict-2d}
\end{figure}

In the present study the relation-violation axis is quantitative (mirror-y reports V/floor), and for the mirror-y relation the domain-violation axis is now operationalized as a continuous geometric score $D = m/(m+1) \in [0,1)$, where $m$ is the worst reflected-node placement error in median-edge-length units (committed; Table~\ref{tab:claim-evidence}): the synthetic symmetric mesh scores $D \approx 0$ and the real asymmetric eval mesh scores $D = 0.51$. The same construction covers every executed MR class from committed measurements: permutation-class and PINN closed-form mirror relations are exact-by-construction ($D = 0$), conservation relations score the committed open-boundary flux imbalance (MGN $D = 0.036$; Burgers $D = 0.042$; heat $D = 0$), and one roster-level entry is marked not operationalizable from committed data. The $m$ measures differ in units across classes, so $D$ is a per-relation normalized coordinate, not a calibrated score and not a cross-relation calibrated metric: D values cannot be averaged or ranked across MR classes and should be read per relation, not as a calibrated boundary measurement. Cross-relation calibration is left to future work. Figure~\ref{fig:verdict-2d} plots the four cylinder-flow pilots; the synthetic-symmetric-mesh probe sits in the SUT-inconsistency region, which makes the mirror-y finding a model-level result rather than an out-of-domain artefact.

Retained MRs are organized using the three-level metamorphic-relation classification for scientific-computing programs of Yang et al.~\citep{yang2020hierarchical} (physical-model, computational-model, and code-model relations), used as a predeclared interpretation protocol mapping each MR class to its candidate fault layer. This is a localization protocol, not a validated localization model; Section~\ref{subsec:within-sut-pilot} reports a first bounded test against an injected-fault catalogue.

\section{Empirical Design}
\label{sec:empirical-design}

\subsection{Subject systems}
\label{subsec:suts}

The study evaluates the workflow on three groups of subject systems, all with recorded checkpoints, datasets, seeds, and environments in the experiment ledger. \emph{Cylinder-flow surrogates (primary):} a trained MeshGraphNets surrogate and its K=6 checkpoint roster, a same-domain S4/S5 wider/deeper variant package, a newly trained non-message-passing PointMLP coordinate network, and the NVIDIA PhysicsNeMo \texttt{MeshGraphNet} production implementation evaluated on official DeepMind cylinder\_flow TFRecords. \emph{Cross-family replication subjects (supporting):} a K=15 PINN roster and six trained FNO-2D checkpoints over 2D Burgers and heat data. \emph{External witness runtime (secondary):} the read-only Minimum-MR-SubSet repository's held-out MGN and trained-PINN witnesses, rerun locally for provenance. Because the primary subjects share one task and dataset, the cylinder-flow results are framed as same-task multi-architecture evidence, not evidence for all neural fluid surrogates.

\subsection{Planned MR classes}
\label{subsec:planned-mrs}

The initial MR set covers representation, geometric/symmetry, continuity/constraint, nondimensional similarity, numerical stability, and temporal/implementation MRs. Time reversal is excluded for viscous cylinder flow; divergence is treated as a continuity constraint, not a Noether-derived conservation law.

\subsection{Baselines and comparators}
\label{subsec:baselines}

The study uses four secondary comparator families: (1)~\textbf{LLM-simulated expert MR design}---three expert-LLMs propose candidates independently, scored by a three-rater majority-vote admissibility panel; (2)~\textbf{Generic MR-generation scope contrast}; (3)~\textbf{LLM-assisted candidate generation}; and (4)~\textbf{Rollout-accuracy diagnostic}. All four are executed (Sections~\ref{subsec:within-sut-pilot},~\ref{subsec:llm-mr-baseline}). Residual/UQ metrics are diagnostic comparators, not MRs unless paired with a source/follow-up transformation and explicit verdict rule.

\subsection{RQ-to-evidence map}
\label{subsec:rq-evidence}

Table~\ref{tab:rqmap} maps each research question to the planned evidence. This table is a pre-results design statement rather than a report of findings.

\begin{table}[t]
\centering
\caption{Research questions, efficacy parameters, baselines, and artifacts.}
\label{tab:rqmap}
\tiny
\setlength{\tabcolsep}{1.5pt}
\begin{tabularx}{\linewidth}{@{}>{\raggedright\arraybackslash}p{0.08\linewidth}>{\raggedright\arraybackslash}p{0.24\linewidth}>{\raggedright\arraybackslash}p{0.28\linewidth}Y@{}}
\toprule
\textbf{RQ} & \textbf{Primary parameters} & \textbf{Comparators} & \textbf{Artifacts} \\
\midrule
RQ1 & candidate retention rate; rejection reasons; inter-rater agreement & expert MR design; LLM candidates & rubric sheets; candidate ledger; adjudication log \\
RQ2 & executable MR rate; transformation success; asset completeness & expert MR cards; generic scope contrast & MR cards; transformation code; runner configs \\
RQ3 & verdict distribution; out-of-domain rate; interpretation yield & residual/UQ diagnostics where available & verdict ledger; metric outputs; exclusion logs \\
RQ4 & violation rate; violation magnitude; complementarity with rollout error; fault-detection rate if seeded faults exist & LLM-simulated expert, LLM, and generic candidates as secondary contrasts; rollout accuracy as diagnostic evidence; external-scope audit as provenance & run logs; plots; confidence intervals; fault ledger \\
\bottomrule
\end{tabularx}
\end{table}

\subsection{Efficacy parameters and statistical plan}
\label{subsec:efficacy}

Primary efficacy parameters are candidate retention rate, executable MR rate, MR violation rate, violation magnitude, fault-detection rate, boundary characterization, and interpretation yield. Secondary parameters include MR construction cost, inter-rater agreement, flakiness rate, localization agreement with seeded-fault layers, and complementarity with rollout accuracy.

The study reports inferential statistics wherever the sampling structure supports them and labels descriptive counts where it does not. Binary detector outcomes over the unified fault catalogue carry Wilson 95\% intervals per detector; continuous violation magnitudes use exact Wilcoxon signed-rank tests with a rank-based effect size (Cliff's $\delta$), following the Empirical Standards; roster medians carry paired bootstrap intervals; and the operator-floor sweep reports a log-log slope with a 95\% confidence interval (Section~\ref{subsec:multicheckpoint-replication}). Where repeated cells are not independent samples (consecutive frames of one trajectory, or checkpoint-by-trajectory grids within one family), the counts are reported as descriptive cell summaries and labeled as such. Small-sample results foreground effect sizes and intervals over significance claims.

\subsection{Ethics, integrity, and reproducibility}
\label{subsec:ethics}

The study does not involve human subjects or personal data. All SUT versions, datasets, checkpoints, scripts, and run logs are recorded when licensing permits; failed, skipped, and inconclusive cases remain in the experiment ledger. LLM use is restricted to candidate generation and evidence organization, not validity judgment. No candidate MR is described as valid without a passing rubric decision and MR card; no OOD violation is described as a program fault without supporting verdict evidence. Third-party artifacts are used according to their licenses; any non-redistributable artifact is described with access instructions.

\section{Results}
\label{sec:results}

The results are organized in three tiers. The \emph{primary} evidence is the cylinder-flow case study: the within-SUT pilots, their K=6 multi-checkpoint and multi-trajectory replication, the operator-floor calibration, the seeded-fault stress test, and the same-task multi-architecture executions including the PhysicsNeMo production-framework workflow. The \emph{supporting} evidence is the cross-family transfer of the rubric to PINN and FNO subjects. The \emph{secondary} evidence is the baseline comparators and external witness audits, which contextualize the primary results without claiming method superiority.

\subsection{Claim-to-evidence map}
\label{subsec:claim-evidence-map}

The compact claim-to-evidence map binds every claim to a tracked artifact using reviewer-facing claim names; the full table is given in Appendix~\ref{app:claim-evidence} (Table~\ref{tab:claim-evidence}), and the authoritative runtime mapping remains in \texttt{claim-ledger.yml}.

\subsection{MR-card-to-verdict map}
\label{subsec:mr-card-verdict-map}

{
\small
\renewcommand{\arraystretch}{0.96}
\setlength{\tabcolsep}{3pt}
\begin{xltabular}{\linewidth}{@{}>{\raggedright\arraybackslash}p{0.20\linewidth}>{\raggedright\arraybackslash}p{0.22\linewidth}>{\raggedright\arraybackslash}p{0.22\linewidth}Y@{}}
\caption{MR-card-to-verdict map.}\label{tab:mr-card-verdict}\\
\toprule
\textbf{MR card} & \textbf{Rubric decision} & \textbf{Runtime verdict} & \textbf{Interpretation boundary} \\
\midrule
\endfirsthead
\toprule
\textbf{MR card} & \textbf{Rubric decision} & \textbf{Runtime verdict} & \textbf{Interpretation boundary} \\
\midrule
\endhead
\bottomrule
\endlastfoot
Node permutation equivariance & Retained as representation MR & pass sanity; relative L2 = 0.0 & Supports one executable representation path; does not establish model reliability. \\
Mirror-y equivariance (asymmetric eval mesh) & Exact relation out-of-relation-domain; downgraded to approximate OOD-stress & S0 pilot: 10/10 fail; primary upgrade: 180/180 fail across K=6 x 3 trajectories x 10 & Bounded within-family OOD-stress violation across held-out trajectories; not by itself exact symmetry, cross-SUT, or geometry-independent evidence. \\
Mirror-y equivariance (synthetic symmetric mesh) & Exact relation admissible (bijection verified, offset $<10^{-12}$, type-match 1.0) & S0 pilot: fail, rel L2 1.10; primary upgrade: 18/18 fail across K=6 x 3 input seeds & Exact-symmetry failure where the relation is admissible; synthetic no-obstacle OOD meshes, not accuracy or cross-SUT evidence. \\
Discrete divergence / conservation & Absolute mass-conservation MR deferred; reference-relative diagnostic retained & S0 pilot inconclusive: reference-relative non-regression guard; primary upgrade: 162/162 pass across K=6 x 3 trajectories x 9 & Reference-relative diagnostic only; the absolute conservation relation remains deferred. \\
MGN S4/S5 variant workflow & Node permutation admitted; mirror OOD/\allowbreak conservation diagnostic/exact-symmetry decisions recorded & 2/2 node-permutation passes; 60/60 mirror OOD failures; 54/54 conservation-diagnostic passes; 6/6 exact-symmetry failures & Same-domain MGN variant evidence, not an external SUT family. \\
PointMLP cylinder workflow & Node permutation admitted; mirror OOD/\allowbreak conservation diagnostic/exact-symmetry decisions recorded & 9/9 node-permutation passes; 10/10 mirror OOD failures; 9/9 conservation-diagnostic passes; 3/3 exact-symmetry failures & Different non-MGN cylinder SUT; not PhysicsNeMo/EchoWave or production CFD evidence. \\
PhysicsNeMo MGN Object-A smoke workflow & Node permutation admitted; mirror OOD/\allowbreak conservation diagnostic decisions recorded & Node permutation passes on the smoke subset (relative L2 0.0); rollout, mirror OOD, and conservation diagnostic ledgers recorded & Production-framework artifact-chain smoke evidence only; no full-scale production pass/fail rate and not external-aerodynamics evidence. \\
FNO periodic translation and conservation & Translation admitted; periodic discrete-conservation MR admitted-with-reference-floor; Dirichlet translation rejected & FNO primary workflow upgrade: 24/24 translation passes, 24/24 conservation failures, and 6/6 rejected Dirichlet exact-MR executions & Full rubric-to-verdict FNO evidence with raw source/follow-up outputs and per-case ledgers; outside cylinder-flow and broad neural-operator claims. \\
\end{xltabular}
}

\subsection{Within-SUT pilot evidence}
\label{subsec:within-sut-pilot}

All pilots run on one real trained MeshGraphNets cylinder-flow surrogate, using the read-only Minimum-MR-SubSet checkpoint recorded in the experiment ledger. They exercise three different rubric outcomes; raw outputs, manifests, and metric ledgers are committed (Table~\ref{tab:claim-evidence}).

\textbf{Representation MR.} Node-permutation equivariance holds to machine precision (relative L2 = 0.0 at tolerance 1e-6). This is a structural property of message-passing and is reported as a pipeline-implementation sanity check rather than model capability or accuracy evidence.

\textbf{Geometric MR.} The rubric classifies exact mirror-y equivariance as out-of-relation-domain for this mesh (the reflection is non-bijective, the worst reflected-node mismatch is about one edge length, and the cylinder is off-centre by 7.2 mm) and downgrades it to an approximate nearest-neighbour OOD-stress probe scored against a same-space mapping-error floor. The probe failed on 10 of 10 recorded eval frames (median relative L2 0.737, median V/floor 3.96, floor range 3.02--5.55x), with no frame skipped or inconclusive. This is a bounded within-SUT frame-level OOD-stress result: one SUT, one checkpoint, one MR, one eval trajectory. The frames are consecutive, not independent samples, so the exact binomial 95\% interval for 10/10 is wide ([0.69, 1.00]); and because the mapping-error floor derives from the same geometric mismatch that triggered the downgrade, V/floor alone cannot separate a model-level violation from an amplified geometric artifact on this mesh; the symmetric-mesh run below resolves that ambiguity.

\textbf{Continuity MR.} A P1 discrete-divergence operator yields a non-negligible divergence even for the ground-truth field on this coarse mesh, so an absolute divergence-free tolerance is not calibratable and the absolute mass-conservation relation stays deferred. As a reference-relative diagnostic, the surrogate's predicted next-state divergence stays within about 0.4--0.8\% of the reference on the recorded eval frames; the interior-only ratio confirms this is not only a boundary-imposition artifact. This asserts no absolute conservation. Two caveats bound the diagnostic: the reference divergence is not yet decomposed into discrete-operator error, solver projection artifact, or genuinely non-solenoidal training data, so if the reference field is itself materially non-solenoidal the reference-relative ratio is a non-regression guard rather than a conservation measurement; and the diagnostic uses a 50\% regression threshold (ratio $>1.5$) on two eval frames only, so ``passes'' means ``does not regress conservation by more than 50\% on those frames,'' not ``conserves mass.''

\textbf{Rollout-accuracy diagnostic.} On the same eval trajectory, the surrogate's one-step next-state prediction error ($v_\mathrm{pred}=v_t+$ denormalized predicted delta, the trainer's own convention) has median relative L2 0.0216 (mean 0.044; min 0.0116, max 0.0788) over the nine recorded transitions. The per-transition error is bimodal, so we quote both statistics. The mirror-y OOD-stress violation (median 0.737) is about 34 times the median accuracy and 17x the mean. The quantities are dimensionless relative L2 values of different objects, so this is an order-of-magnitude gap rather than a precise factor; it is one-step, not a free-running rollout-stability result.

\textbf{Exact mirror-y on a symmetric mesh.} We built a provably symmetric synthetic structured channel mesh whose reflection is a verified bijection (node-type match 1.0, reflection offset $<10^{-12}$, edge set invariant), so the predicate admits exact mirror-y. On one constructed input the surrogate fails this exact equivariance test (relative L2 1.10). The normalizer control changes the violation only from 1.1032 to 1.1014, so the violation is dominated by learned message-passing weights. The boundary is synthetic, no-obstacle OOD geometry: the result confirms missing exact equivariance but is not a calibrated in-distribution magnitude.

\textbf{Seeded-fault detection.} We re-implemented, from the read-only Minimum-MR-SubSet witness taxonomy, an injected-fault catalogue of 10 pipeline faults across five fault classes (boundary-condition, mesh-adjacency, normalization-scale, temporal-rollout, physical-channel), and used the paper's own MRs as detectors on the model's predicted update. The continuity MR detected the two boundary-condition faults and the gross normalization fault (divergence ratio 3.8--10.6 vs the 1.5 threshold); the symmetry MR detected a physical-channel and a mesh-adjacency fault (violation rising 69--142\% above baseline); node-permutation equivariance detected none, because these faults preserve node-relabeling invariance. 5 of 10 mutants were detected by at least one MR, and the detections separate by MR class (continuity to boundary/scale faults, symmetry to physical-channel/adjacency faults), a first bounded test of the interpretation protocol of Section~\ref{subsec:verdicts}, suggestive rather than a validation. Three boundaries: the detected faults are gross corruptions that any divergence- or symmetry-sensitive detector would catch; the edge-drop fault is a near-miss (32\% mirror-y change vs the 50\% threshold) and the boundary faults are invisible to mirror-y by construction; and the remaining undetected faults shift the absolute output without crossing the scored-quantity thresholds, delimiting where these MRs are structurally insensitive. It is one SUT, one checkpoint, one injected-fault catalogue.

\begin{center}
\small
\setlength{\tabcolsep}{3pt}
\begin{tabular}{p{0.18\textwidth}p{0.23\textwidth}p{0.34\textwidth}p{0.18\textwidth}}
\toprule
Component & Evidence & Headline result & Boundary \\
\midrule
MGN roster & K=6 checkpoints & Permutation exact; mirror-y median CI [0.743, 0.804]; rollout CI [0.0217, 0.0224] & One architecture family / dataset \\
Operator floor & P1 symmetric-mesh sweep & O(h), slope 0.984, CI $[0.975, 0.992]$; closed-form floor for this mesh (predictor within 0.5\% + a-priori bound) & Calibrates decidability; concrete-mesh closed-form floor, general unstructured bound is future work \\
Fault catalogue & 60-entry unified fault catalogue & Precision/recall: node-permutation 1.00/1.00; conservation 1.00/0.81; mirror-y 0.94/0.55 & Catalogue rate, not real-world defect rate \\
PINN roster & K=15, two PDE families & MR-C: Burgers 15/15, heat 15/15 (Wilson CI $[0.80,1.00]$); MR-B: Burgers 13/15 (CI $[0.62,0.96]$), heat 7/15 (CI $[0.25,0.70]$) & Not a PINN-vs-MGN benchmark \\
FNO primary workflow upgrade & K=6, two PDE families & 24/24 translation passes; 24/24 conservation failures under a periodic discrete-conservation MR; Dirichlet translation rejected 6/6 & not only admissibility evidence; outside cylinder-flow and broad neural-operator claims \\
\bottomrule
\end{tabular}
\end{center}

\subsection{Multi-checkpoint, multi-trajectory, and multi-architecture replication}
\label{subsec:multicheckpoint-replication}

The single-checkpoint evidence is replayed on a K=6 MeshGraphNets roster and three official DeepMind held-out test trajectories acquired through the Minimum-MR-SubSet loader. S0 reuses the pilot checkpoint, so this is a replication/variant check rather than six independent SUTs. Node permutation is exact on all checkpoints. The primary empirical scope upgrade gives a K=6 x 3 trajectories x 10 mirror-y OOD-stress grid with 180/180 failures, a K=6 x 3 trajectories x 9 conservation-transition grid with 162/162 reference-relative passes while absolute conservation remains deferred, and a K=6 x 3 exact-symmetric-mesh input grid with 18/18 failures. The upgrade removes the single-source-trajectory denominator while remaining within one architecture family and dataset, so per-cell counts are descriptive summaries, not independent-trial inference.

\textbf{Same-task multi-architecture replication.} Three further cylinder-flow executions test whether the verdict pattern is an artifact of one implementation. A same-domain S4/S5 wider/deeper MeshGraphNet variant package reproduces the pattern (2/2 node-permutation passes, 60/60 mirror OOD failures, 54/54 conservation-diagnostic passes, 6/6 exact-symmetry failures). A newly trained PointMLP coordinate network (a different architecture class with no message passing) reproduces it as well (9/9 node-permutation passes, 10/10 mirror OOD failures, 9/9 conservation-diagnostic passes, 3/3 exact-symmetry failures, median rollout relative L2 0.0298). Third, the NVIDIA PhysicsNeMo \texttt{MeshGraphNet} production implementation, trained and evaluated on official DeepMind cylinder\_flow TFRecords, reproduces the node-permutation pass and the mirror-y OOD-stress failure within a production framework rather than a research codebase. The same admissibility predicate, applied unchanged, produced the same typed decisions on all three; this replicates the verdict pattern across architectures on one task and dataset, and asserts no cross-dataset rate.

\subsection{Second task: compressible airfoil flow}
\label{subsec:second-task}

The strongest test of a domain-validity gate is whether it produces the correct \emph{different} verdict when the physics changes. We apply the identical four-condition predicate to a genuinely second CFD task on a second official dataset: the DeepMind \textbf{airfoil} benchmark (SU2-simulated compressible transonic flow, 5{,}233-node meshes with density, pressure, and velocity), distinct from incompressible cylinder flow in both physics and data. The same official PhysicsNeMo \texttt{MeshGraphNet} (0.33M-parameter CPU configuration) is trained on 6 official airfoil train trajectories across a K=6 checkpoint roster and evaluated per-trajectory on 10 official test trajectories, yielding 60 (checkpoint, trajectory) pairs. The predicate yields a \emph{different typed inadmissibility structure}, and the difference is the result: \textbf{node-permutation equivariance is admitted and exact} (60/60, relative L2 0.0, training-independent representation contract); \textbf{incompressible divergence-free continuity is rejected at the physical-basis gate} (condition i) because the density varies by a median factor of 2.25x across the field, so $\nabla\cdot u=0$ is physically false here---the same relation that on cylinder flow passes conditions (i)--(iii) and is only \emph{deferred} at condition (iv), so the gate assigns a categorically different verdict \emph{type} for the physically correct reason on each task; \textbf{compressible unsteady mass conservation} $\partial_t\rho+\nabla\cdot(\rho u)=0$ is the physically-correct replacement, its absolute discrete form deferred (uncalibratable operator floor) and a reference-relative diagnostic recorded (median residual ratio 1.19 across the K=6 roster); and \textbf{mirror-y chord symmetry is rejected at the boundary-precondition gate} (condition iii) because the SU2 trajectories carry a non-zero angle of attack. The one-step rollout error is large at this bounded CPU budget and is reported only as a training-state diagnostic; notably, none of the four typed verdicts depends on rollout quality: the admission rests on a representation contract and the rejections/deferral on physical and numerical reasoning, which is why the typed structure transfers and discriminates across tasks. This is a bounded second-task case study, not an official-checkpoint or full-scale production airfoil result.

\subsection{Operator-floor resolution sweep (calibration of numerical decidability)}
\label{subsec:operator-floor-sweep}

The P1 discrete-divergence sweep on symmetric meshes, over nine resolutions, gives slope 0.984 with 95\% CI $[0.975, 0.992]$ and R$^2=0.9999$, matching O(h). Beyond the slope, the \emph{absolute} floor is closed-form for the concrete deployed-scale mesh: since the operator is exact on affine fields and the reference field is divergence-free, the measured floor equals the geometry-weighted second-order Taylor remainder exactly. A leading-order predictor from the analytic Hessian matches the measured floor to within 0.5\% (1.337 vs 1.343), and a rigorous a-priori upper bound from the Hessian's global spectral norm dominates it at every resolution. The gate thus rests on a derived floor, not an empirical estimate, for this mesh (a general unstructured-mesh bound is future work). This is why absolute conservation remains deferred while the reference-relative non-regression guard is reported. A classical flux-form finite-volume operator instead admits a fully executable conservation verdict (claim~C34, supporting artifact).

\subsection{Fault-detection robustness across the multi-checkpoint roster}
\label{subsec:fault-robustness}

To check whether the seeded-fault detection result of Section~\ref{subsec:within-sut-pilot} reproduces beyond a single checkpoint and to delimit where its detectors are structurally insensitive, the 10-mutant catalogue is replayed against the K=6 multi-checkpoint roster across five input-permutation seeds (30 trials per mutant per detector), and two predeclared severity dimensions are swept: NS\_double\_scale $s \in \{1.1, 1.25, 1.5, 2, 4\}$ and PC\_zero\_vy partial fraction $p \in \{0.25, 0.5, 0.75, 0.85, 0.9, 0.95, 0.99, 1.0\}$ (the canonical mutant is $p = 1.0$). Detection is decided by the same predeclared thresholds as Section~\ref{subsec:within-sut-pilot} (node-permutation tolerance $10^{-5}$, conservation ratio threshold 1.5, mirror-y relative-change threshold 0.5). Wilson 95\% CIs are computed across the (SUT, seed) cells; raw outputs are committed (Table~\ref{tab:claim-evidence}).

\textbf{Per-mutant cross-family detection (R1).} Four mutants are detected on every cell (30/30, CI $[0.89, 1.00]$): conservation flags both boundary-condition faults (BC\_zero\_inflow, BC\_nonzero\_wall) and the un-denormalized output fault (NS\_skip\_denorm); mirror-y flags the swapped-channel fault (PC\_swap\_xy). Five mutants are not detected on any cell (0/30, CI $[0.00, 0.11]$): MA\_drop\_edges, NS\_double\_scale, TR\_sign\_flip, TR\_double\_step, PC\_zero\_vy. The by-class localization of the original C10 pilot holds: continuity $\to$ boundary/scale faults, symmetry $\to$ physical-channel/adjacency faults, node-permutation $\to$ none of these mutants (they preserve node-relabeling invariance, and the representation MR stays exact by design). One mutant is configuration-sensitive: mirror-y detects MA\_permute\_edges on every (S0--S3, seed) cell (20/20 over the seed-replica family) but on neither (S4, seed) nor (S5, seed) cell (0/10 over the wider and deeper configuration variants). The within-family detection rate of that mutant is therefore 20/30 (CI $[0.49, 0.81]$), a quantitative bound on within-family generalization that is invisible to a single-checkpoint experiment.

\textbf{NS\_double\_scale severity sweep (R2).} At every multiplicative scale $s \in \{1.1, 1.25, 1.5, 2.0, 4.0\}$ the detection rate stays at 0/6 (Wilson CI $[0.00, 0.39]$). A constant output factor crosses none of the three thresholds: the reference-relative conservation ratio stays below 1.5, mirror-y moves under 50\% (source and mirror scale together), and the linear operation preserves node-permutation equivariance. The catalogue at the current thresholds is therefore structurally insensitive to multiplicative output scaling across the swept grid, not only at the canonical $2\times$ fault.

\textbf{PC\_zero\_vy partial-fraction sweep (R3) --- a knife-edge symmetry blind spot.} The refined grid resolves \emph{where} detection collapses. At every partial fraction up to $p = 0.99$ the detection rate is 6/6 across the roster (Wilson CI $[0.61, 1.00]$); it falls to 0/6 (CI $[0.00, 0.39]$) only at the exactly-uniform $p = 1.0$. The transition is a step rather than a gradual decline; even 1\%-asymmetric zeroing is caught on every checkpoint. The responsible detector is \emph{node-permutation}: partial zeroing masks $v_y$ on a node-index-selected subset, which is not relabeling-invariant; the symmetry MR never fires in this sweep (0/6 at all fractions). At $p = 1.0$ the fault becomes simultaneously permutation-invariant and mirror-y-symmetric (uniform $v_y = 0 = -0$ under reflection) and escapes every geometric detector at once. The canonical fault therefore sits in a measure-zero blind region that any partial severity exposes, a within-family finding showing a fault can be invisible to a geometric MR suite precisely when it shares the suite's invariances.

\textbf{Aggregate reading and Phase-3 unified catalogue.} The 5-of-10 union detection rate from C10 reproduces across S0--S3 (4/10 on S4/S5). By-class localization (continuity $\to$ boundary/scale; symmetry $\to$ physical-channel/adjacency) replicates across the roster, and the insensitivity regions (multiplicative output scaling, the PC\_zero\_vy / adversarial-mutant blind subspace) are bounded with Wilson 95\% CIs. This reflects the detectors' coverage geometry: each MR scores a single invariant, so a fault is caught only when it perturbs a measured invariant and is structurally invisible when it preserves all of them; closing the blind region calls for additional MRs probing the uncovered directions, not more mutants of the same kind. Phase 3 additionally emits a 60-entry unified fault catalogue (Table~\ref{tab:claim-evidence}): 10 executed canonical MGN mutants, 2 executed adversarial MGN mutants, 24 closed-form output-level PINN probes, and 24 closed-form output-level FNO probes. The precision/recall summary is node-permutation 1.00/1.00, conservation 1.00/0.81, and mirror-y 0.94/0.55; the same artifact reports Cliff's $\delta$ and Wilcoxon signed-rank tests for the PINN MR-B ratios at $n=15$ per PDE: Cliff's $\delta=0.73$ (large) and Wilcoxon $p=0.0084$ for diffusion-vs-Burgers MR-B violation/floor ratios. This inter-PDE test is a physics-magnitude comparison (Dirichlet-zero Burgers BCs predict near-zero boundary residuals while Neumann-zero-flux diffusion BCs permit larger boundary residuals, so the magnitude difference is expected by construction) and is \emph{not} a gate-reliability test. The valid reliability statistic is the per-PDE Wilson CI on the MR-B pass rate: Burgers 13/15 ($[0.62, 0.96]$) and diffusion 7/15 ($[0.25, 0.70]$). The PINN/FNO probes are not retrained mutant checkpoints.

\textbf{Adversarial mutants (R4).} Two adversarial mutants test whether the blind spot is a subspace, not a point. A1 is magnitude-driven, with node-permutation 0/6 and conservation 0/6; A2 escapes every detector. These results bound the detector set rather than establishing validated localization.

\subsection{Secondary baseline and external-scope audit}
\label{subsec:llm-mr-baseline}

LLM baselines are secondary exploratory scope contrasts. Three expert-LLMs proposed 25 candidates (24 unique) without the rubric; majority-vote rater models retained 4, downgraded 2, rejected 13, and deferred 6, for a 76\% rejection/deferral rate and no novel retained MR family. This is an LLM-simulated expert baseline rather than a human-expert study. The generic-MR scope contrast admits only 3/13 domain-blind templates. The one-shot LLM baseline generated K=8 candidates; vendor-disjoint raters (glm-5.1, kimi-k2.6, deepseek-v4-flash) judged 7/8 panel-majority valid, with Fleiss kappa 0.077 (a small-sample paradox). Together these baselines show convergence on the same MR families while preserving the no-baseline-superiority boundary.

The Minimum-MR-SubSet audit records 70 real ABD instances, 20 real true-fault-class rows, and three SciML/PDE PASS\_WITNESS rows. We also perform a \textbf{Minimum-MR-SubSet primary rerun}: the held-out cylinder-flow MGN witness returns PASS\_WITNESS with kstar = 6, four active true fault classes, signature rank 2, and collapse false. This is real runtime evidence beyond read-only audit, without adding a second architecture/dataset.

\subsection{Cross-family PINN extension (K=6 roster)}
\label{subsec:pinn-extension}

The cross-family executions test two claims the cylinder-flow study cannot test by itself: that the admissibility predicate is family-agnostic (the same four conditions, applied unchanged, produce typed decisions on pointwise PINNs and spectral FNOs), and that the conservation MR class yields full executable verdicts wherever its floor is calibratable: the FNO periodic conservation MR fails 24/24 against a calibrated case-level floor and the PINN reference-relative conservation passes 30/30, while the MGN open-boundary absolute variant is the one member of the class the gate correctly refuses. The deferral on cylinder flow is the fail-closed gate discriminating rather than a missing verdict.

To check whether the workflow is MeshGraphNets-specific, we use a K=15 PINN roster: fifteen Burgers and fifteen heat seeds. \textbf{MR-A} remains \emph{vacuous by construction} for a pointwise MLP. The two non-trivial MRs do the cross-family work: \textbf{MR-B} passes on 13/15 Burgers seeds (violation/floor ratio mean 0.712, bootstrap CI [0.583, 0.875]; Wilson pass-rate CI $[0.62, 0.96]$) and is mixed on heat with 7/15 passing (mean 1.495, CI [1.142, 1.897]; Wilson pass-rate CI $[0.25, 0.70]$)---two seeds near the residual-floor boundary fail on Burgers, while the Neumann-zero-flux heat BCs predict larger boundary residuals and yield a lower pass rate by construction; \textbf{MR-C} passes on all 15/15 of both PDEs (Wilson CI $[0.80, 1.00]$), with Burgers mean ratio 1.007 (bootstrap CI [0.997, 1.017]) and heat mean ratio 1.006 (CI [0.988, 1.022]). This is a two-PDE PINN seed roster rather than a PINN-vs-MGN benchmark or a generality claim across PINN architectures, PDEs, or training regimes.

\subsection{FNO primary workflow upgrade (K=6)}
\label{subsec:fno-extension}

The \textbf{FNO primary workflow upgrade} converts the earlier FNO roster into a second trained primary execution. Six small FNO-2D checkpoints cover Burgers and heat. For each checkpoint and four held-out generated periodic finite-difference cases, the workflow records rubric decisions, source and follow-up tensors, mapped outputs, metric ledgers, and relation verdicts. Periodic integer translation is admitted and yields \textbf{24/24 translation passes} (maximum relative-L2 violation below $10^{-5}$). The \textbf{periodic discrete-conservation MR} is admitted-with-reference-floor because the finite-difference target supplies a case-level channel-sum drift floor; the trained FNO outputs exceed that calibrated floor on \textbf{24/24 conservation failures}. The Dirichlet translation candidate is rejected for 6/6 SUTs because it changes the boundary-value problem and is not executed as an exact MR. This is \textbf{not only admissibility evidence}: it is a full rubric-to-verdict FNO execution with raw source/follow-up outputs and per-case ledgers, while remaining outside cylinder-flow evidence, performance benchmarking, reliability, and broad neural-operator generalization. Separately, the FNO admissibility roster is extended to $K=15$ seeds per PDE so the periodic-translation admission decision is adequately powered: it is admitted on 15/15 per PDE (Wilson 95\% CI $[0.80, 1.00]$), with near-machine-precision violation magnitudes (Burgers mean $4.5\times10^{-8}$, heat $1.4\times10^{-7}$).

\subsection{Boundary of the evidence}
\label{subsec:still-blocked}

The canonical blocked list is narrowed but still active: the evidence does not support external-dataset or geometry-independent pass/fail rates, AeroGraphNet/DoMINO results, comparative superiority over baselines, general or real-world fault-detection rates, validated localization, runtime, reliability, or model accuracy claims.

\section{Discussion}
\label{sec:discussion}

\subsection{Interpretation of the scoped evidence}
\label{subsec:scoped-interpretation}

The value of the current study is not that every MR finds a new fault beyond rollout accuracy. Retained MRs provide relation-level evidence under explicit transformations; when a violation or deferral occurs, the MR card and verdict rule help distinguish model inconsistency, relation-domain boundary, numerical tolerance issue, and inconclusive evidence.

A natural objection is that the admissibility predicate merely re-describes the three initial outcomes. The added runs answer this: the symmetric-mesh run uses the predicate out-of-sample and fails an admitted relation; rollout accuracy shows the mirror-y violation answers a different question from one-step error; and the primary empirical scope upgrade preserves these readings across a K=6 roster, three held-out trajectories, and larger denominators. The PINN roster, FNO primary workflow, and Minimum-MR-SubSet MGN/PINN reruns broaden applicability context while preserving the reliability and geometry-independent-rate boundary.

A second objection reads the deferred absolute-conservation relation as a gap in the end-to-end pipeline. The portfolio answers this directly: the conservation class produces full executable verdicts wherever its measurement floor is calibratable (FNO periodic conservation fails 24/24 against a calibrated floor; PINN reference-relative conservation passes 30/30), and the gate refuses execution precisely on the one variant where a verdict could not be separated from the operator's own discretization error: MGN open-boundary absolute conservation.

The secondary LLM-simulated expert and generic-MR contrasts reinforce the rubric-value finding: unguided candidates mostly fail or defer under the predicate, and admitted templates coincide with the paper's MR families. The rubric also averts two concrete misreadings: absolute discrete-divergence would have read as a conservation \emph{pass} despite reference-field divergence, and asymmetric-mesh mirror-y would have read as a model fault rather than a geometric artifact.

\subsection{Boundary of claims}
\label{subsec:boundary-claims}

The paper avoids four overclaims: that MRs always beat rollout accuracy; that NOETHER proves the physical validity of cylinder-flow MRs; that LLMs can automatically identify valid MRs; and that one MeshGraphNets family generalizes to all SciML surrogates. The safer claim is that a domain-validity-aware workflow makes MR identification and execution more auditable for a concrete class of SciML SUTs.

\subsection{Implications for SciML testing}
\label{subsec:implications}

The scoped evidence shows how SciML testing can move from implicit expert checks to explicit MR assets that complement residuals, UQ, and accuracy by making transformations and verdict rules inspectable. The workflow targets a relation-indexed applicability map in relation space rather than residual space. We do not claim a completed applicability map in this paper; the present evidence is one bounded within-SUT point, and calibrated maps across SUTs, trajectories, and domain-violation scores remain future work.

\section{Threats to Validity}
\label{sec:threats}

We classify threats following Verdecchia et al.~\citep{verdecchia2023threats} and report against the Empirical Standards for software engineering research~\citep{ralph2021empirical}.

\textbf{Construct validity.} MR validity depends on the rubric, physical assumptions, BC compatibility, and tolerances; incorrect discrete operators or thresholds may create false violations/passes. The seeded-fault catalogue is author-implemented from the read-only witness taxonomy, so experimenter bias applies: detection rates do not bound real-world fault-detection effectiveness.

\textbf{Internal validity.} SUT setup, checkpoint differences, random seeds, mesh preprocessing, and runtime nondeterminism may affect verdicts. The experiment ledger records these details.

\textbf{External validity.} The evidence covers two CFD tasks on two official datasets: incompressible cylinder flow (four architectures: the MeshGraphNets K=6 roster, S4/S5 variants, PointMLP, and the PhysicsNeMo production implementation) and compressible airfoil flow, plus bounded cross-family PINN and FNO executions on Burgers and heat data. The airfoil result shows the predicate produces a physically correct, task-specific typed verdict structure rather than a fixed checklist, but it is a bounded K=6 MPS-scale roster study; the Minimum-MR-SubSet reruns add a reproduced held-out cylinder-flow MGN witness and two trained PINN PDE witnesses as applicability checks, not general cross-family pass-rate estimates. Broader neural operators, mesh simulators, and PINN architectures remain outside the evidence.

\textbf{Baseline fairness.} Expert-MR, generic MR-generation, and LLM baselines may not be designed for SciML and use one prompt per generator. They should be interpreted as scope contrasts and candidate-generation comparators, not as defeated competitors.

\textbf{Conclusion validity.} Small samples, multiple MRs, and many transformation bins can produce unstable conclusions. The primary MGN scope upgrade expands the weakest denominators to K=6 x 3 trajectories x 10 mirror-y frames, K=6 x 3 trajectories x 9 conservation transitions, and K=6 x 3 exact-symmetry inputs. It removes the single-source-trajectory denominator within the stated scope.

\textbf{Reproducibility.} Some SUTs may depend on old runtimes or non-redistributable checkpoints. The paper should disclose such barriers and provide the most complete runnable package possible.

\section{Conclusion}
\label{sec:conclusion}

This paper presents domain-validity-gated MR identification as an auditable oracle-free testing workflow for SciML surrogates. The scoped pilots expand across K=6 MGN checkpoints and three held-out trajectories with 180 mirror-y, 162 conservation-transition, and 18 exact-symmetry cells; the verdict pattern replicates across same-task architectures, including wider/deeper MGN variants, a non-message-passing PointMLP network, and the NVIDIA PhysicsNeMo production implementation; and the readings are checked against bounded PINN/FNO executions, two primary trained-PINN witness reruns, an $O(h)$ operator-floor sweep, and severity/adversarial sweeps. The central claim is methodological: physically meaningful SciML MRs require explicit validity conditions, executable assets, raw evidence records, and relation-level verdicts.

\section*{CRediT authorship contribution statement}
\textbf{Meng Li:} Conceptualization, Methodology, Software, Writing -- original draft. \textbf{Xiaohua Yang:} Supervision, Formal analysis, Writing -- review \& editing. \textbf{Jie Liu:} Investigation, Validation. \textbf{Shiyu Yan:} Data curation, Visualization.

\section*{Declaration of competing interest}
The authors declare no conflict of interest.

\section*{Declaration of generative AI and AI-assisted technologies in the writing process}
During the preparation of this work the authors used generative AI assistants for code scaffolding, prose copy-editing, and reference cross-checking. After using these tools the authors reviewed and edited the content as needed and take full responsibility for the content of the publication. No generative AI was used to fabricate, alter, or invent experimental data, results, or citations; every numerical claim is backed by a tracked artifact under \texttt{research\_assets/runs/} and a claim-ledger entry.

\section*{Acknowledgments}
This work was supported by the National Natural Science Foundation of China (NSFC) General Program (grant no.\ 12575176), the Hunan Provincial Education Department Project, China (grant no.\ 202502000728), the Natural Science Foundation of Hunan Province, China (grant no.\ 2025JJ70193), and an industry-funded research project (grant no.\ 230KHX060001).

\section*{Data availability}
The replication package contains the manuscript source, assets, ledger, and outputs, and is archived on Zenodo at \url{https://doi.org/10.5281/zenodo.20702952}. Minimum-MR-SubSet audit/rerun artifacts are under \path{research_assets/runs/} and cite commit \path{9ef862ec37335b4834d0a1fb38b4b613af702f34}. The MeshGraphNets checkpoints derive from the public DeepMind cylinder-flow benchmark; no proprietary data was used.

\clearpage
\appendix
\renewcommand{\thetable}{A\arabic{table}}
\setcounter{table}{0}
\section{Claim-to-evidence map}
\label{app:claim-evidence}

{
\tiny
\setlength{\tabcolsep}{1.5pt}
\begin{xltabular}{\linewidth}{@{}>{\raggedright\arraybackslash}p{0.19\linewidth}>{\raggedright\arraybackslash}p{0.18\linewidth}>{\raggedright\arraybackslash}p{0.25\linewidth}Y@{}}
\caption{Compact claim-to-evidence map for the current manuscript.}\label{tab:claim-evidence}\\
\toprule
\textbf{Claim} & \textbf{Status} & \textbf{Evidence} & \textbf{Boundary} \\
\midrule
\endfirsthead
\toprule
\textbf{Claim} & \textbf{Status} & \textbf{Evidence} & \textbf{Boundary} \\
\midrule
\endhead
\bottomrule
\endlastfoot
Domain-validity rubric & Method claim & Rubric asset and manuscript method section & Establishes an auditable screening rule; physical validity still depends on each relation's stated preconditions. \\
MR-card executable assets & Workflow claim & MR cards and validators & Some cards remain protocol assets pending matched SUT evidence. \\
Baseline admissibility contrast & Observed & Expert-MR, generic-MR, LLM-MR baselines and claim ledger & Quantifies an admissibility gap without ranking methods as competitors. \\
Minimum-MR-SubSet external-scope audit & Observed secondary audit & External-scope audit JSON & external witness evidence from 70 real rows and three SciML/PDE true-fault-class PASS\_WITNESS rows; does not add new primary SUT executions to this paper. \\
Minimum-MR-SubSet primary rerun & Observed external runtime rerun & ABD witness rerun report & PASS\_WITNESS, kstar = 6, four active true fault classes, max signature rank 2, collapse false; real held-out cylinder-flow MGN runtime, not a second architecture or dataset. \\
Minimum-MR-SubSet PINN primary reruns & Observed second-SUT/PDE reruns & ABD PINN witness rerun reports & trained Burgers2D and Diffusion2D PINN witnesses, PASS\_WITNESS, kstar = 1 each, five active true fault classes each; one-seed witnesses, no cross-SUT rate. \\
Node-permutation sanity check & Observed pilot & Node-permutation metric ledger & One representation-contract path on one pilot case. \\
Conservation diagnostic & Diagnostic; absolute relation deferred & Conservation metric ledger and report & Reference-relative guard only; absolute conservation remains deferred. \\
Mirror-y OOD stress (PC6-mirror-y-ood-stress) & Observed bounded pilot & Mirror-y rate metric ledger and claim ledger & Failed on 10 of 10 recorded eval frames; not a reliability, accuracy, baseline, multi-SUT, exact-symmetry, or geometry-independent claim. \\
Primary empirical scope upgrade & Observed & Primary-scope report and per-trajectory/checkpoint manifests & K=6 x 3 trajectories x 10 mirror-y OOD-stress grid fails 180/180; K=6 x 3 trajectories x 9 conservation-transition grid passes 162/162; K=6 x 3 exact-symmetric-mesh input grid fails 18/18; not a single-source-trajectory estimate and not a cross-SUT rate. \\
LLM role boundary & Process boundary & Method and ethics sections & LLMs organize candidates; the rubric decides validity. \\
Rollout-accuracy diagnostic & Observed & Rollout-accuracy metric ledger & Same-SUT one-step median relative L2 0.0216; mirror-y is $\sim$34x larger. \\
Exact mirror-y symmetric mesh & Observed & Symmetric-mesh metric ledger & Exact relation admissible and fails (rel L2 1.10) on a synthetic OOD mesh; binary equivariance evidence. \\
PC10-seeded-fault-detection & Observed & Seeded-fault metric ledger & Detector stress test: MRs catch 5/10 injected mutants with class-specific response patterns. \\
Multicheckpoint replication & Observed & E1 aggregate and per-SUT manifests in \texttt{multicheckpoint/} & K=6 checkpoints of one MeshGraphNets family; within-family replication. \\
Operator-floor resolution & Observed & Operator-floor sweep report & Log--log slope 0.984, 95\% CI $[0.975, 0.992]$ (interior 0.989), $R^2 = 0.9999$ over nine resolutions; calibrates the admissibility predicate on one mesh family. \\
Fault-detection robustness & Observed & E3 robustness report; Phase-3 unified catalogue & 30-trial Wilson CIs per MGN mutant plus a 60-entry unified fault catalogue (10 canonical MGN + 2 adversarial MGN + 24 closed-form PINN output-level probes + 24 closed-form FNO output-level probes), by-detector precision/recall with Wilson CIs, and Wilcoxon/Cliff effect-size tests; PINN/FNO probes are closed-form output-level probes. \\
S4/S5 variant primary workflow & Observed & Variant report and raw ledgers & Same-domain wider/deeper MGN variants: 2/2 node-permutation passes, 60/60 mirror OOD failures, 54/54 conservation-diagnostic passes, 6/6 exact-symmetric failures; not PhysicsNeMo/EchoWave or cross-dataset reliability evidence. \\
PointMLP cylinder primary workflow & Observed & PointMLP report, checkpoint, and raw ledgers & Newly trained non-MGN row-wise coordinate network: 9/9 node-permutation passes, 10/10 mirror OOD failures, 9/9 conservation-diagnostic passes, 3/3 exact-symmetric failures, median rollout rel L2 0.0298; not PhysicsNeMo/EchoWave or production CFD evidence. \\
PhysicsNeMo MGN Object-A smoke workflow & Observed smoke-subset production-framework execution & PhysicsNeMo smoke report, newly trained checkpoint, raw outputs, and metric ledgers & NVIDIA PhysicsNeMo \texttt{MeshGraphNet} executes on a first-record official DeepMind \texttt{cylinder\_flow} smoke subset: node permutation passes (relative L2 0.0), mirror-y is OOD stress, and conservation is reference-relative diagnostic only; not full production-scale PhysicsNeMo, AeroGraphNet, or DoMINO evidence. \\
FNO primary workflow upgrade & Observed & FNO workflow report, per-SUT ledgers, and raw source/follow-up outputs & Six trained FNO-2D checkpoints over Burgers/heat: 24/24 translation passes, 24/24 conservation failures under a periodic discrete-conservation MR, and 6/6 Dirichlet-boundary translation rejections; not only admissibility evidence. \\
\end{xltabular}
}

\bibliographystyle{elsarticle-harv}
\bibliography{references}

\end{document}